\documentclass[prb,aps,twocolumn,superscriptaddress,floatfix,citeautoscript,showpacs]{revtex4}
\usepackage{graphicx,rotating,subfigure,amsmath,amsfonts,amssymb,delarray,color}
\usepackage{dsfont}
\usepackage[T1]{fontenc}
\usepackage{multirow}
\usepackage{comment}

\newcommand{\e}{\text{e}}
\newcommand{\im}{\text{i}}

\newcommand{\tr}{\textrm{tr}}

\def\12{\frac{1}{2}}

\begin{document}
\bibliographystyle{apsrev}

\title{Localization of fermions in coupled chains with identical disorder}
\author{Y. Zhao}
\affiliation{Asia Pacific Center for Theoretical Physics (APCTP), Pohang, Gyeongsangbuk-do, 790-330, Korea}
\affiliation{Department of Applied Physics, School of Science, Northwestern Polytechnical University, Xi'an 710072, China}
\author{S. Ahmed} 
\affiliation{Department of Physics and Astronomy, University of Manitoba, Winnipeg R3T 2N2, Canada}
\author{J. Sirker} 
\affiliation{Department of Physics and Astronomy, University of Manitoba, Winnipeg R3T 2N2, Canada}

\date{\today}

\begin{abstract}
We study fermionic ladders with identical disorder along the leg
direction. Following recent experiments we focus, in particular, on
how an initial occupation imbalance evolves in time. By considering
different initial states and different ladder geometries we conclude
that in generic cases interchain coupling leads to a destruction of
the imbalance over time, both for Anderson and for many-body localized
systems.
\end{abstract}


\maketitle
\section{Introduction}
\label{Intro}
\begin{figure}
  \includegraphics*[width=0.99\columnwidth]{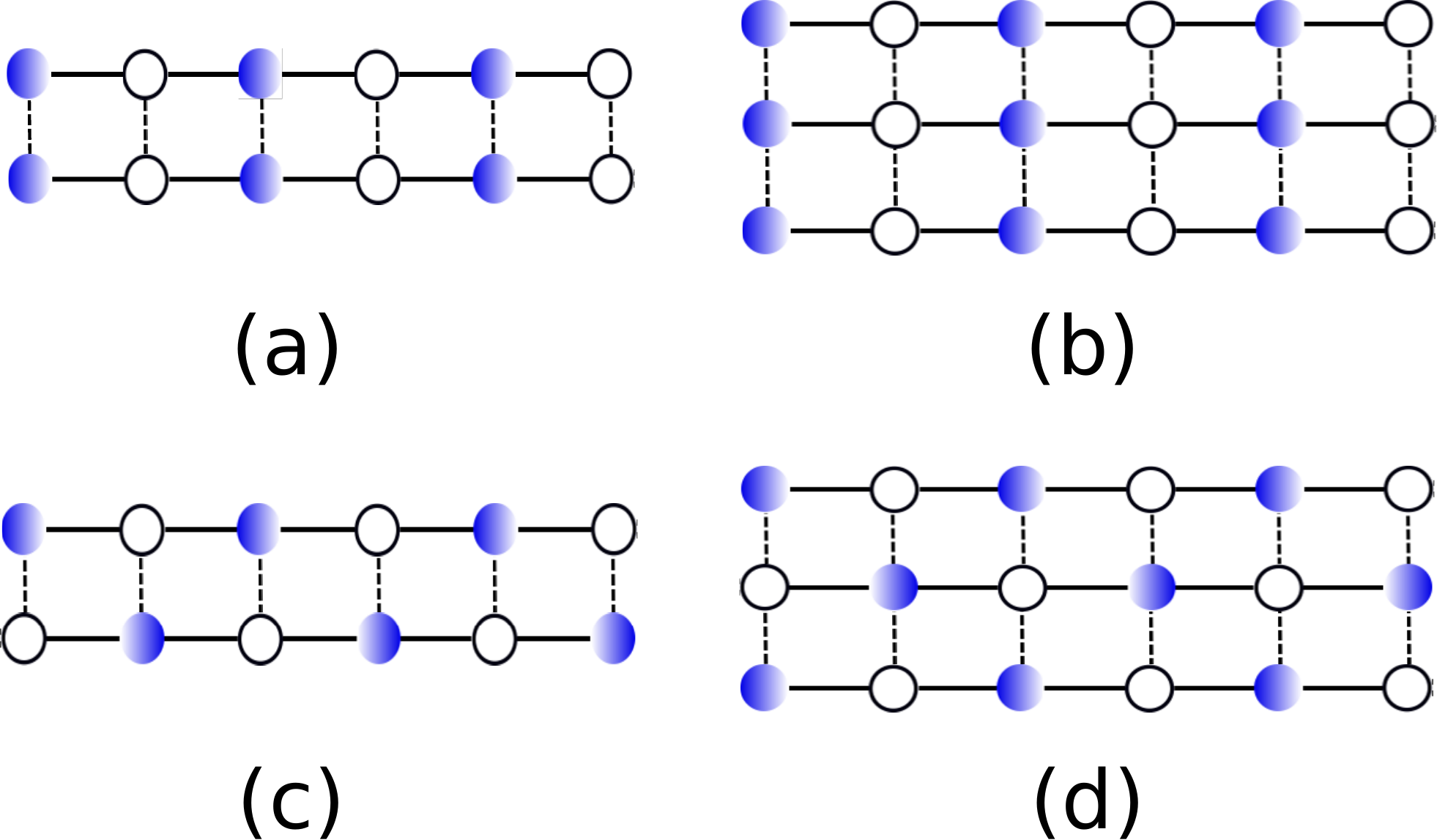}
\caption{(a) and (b) show the rung occupied initial state on two- and three-leg
  ladders while (c) and (d) depict the diagonally occupied initial
  state. }
\label{Fig1}
\end{figure}

It is known for more than fifty years that disorder in one- and
two-dimensional tight-binding models of non-interacting fermions with
sufficiently fast decaying hopping amplitudes always leads to
localization
\cite{Anderson,AbrahamsAnderson,EdwardsThouless,AndersonLocalization,KramerMacKinnon}. In recent years, localization phenomena in interacting 
low-dimensional tight-binding models have attracted renewed attention
\cite{AleinerAltshuler,ZnidaricProsen,PalHuse,Imbrie2016,NandkishoreHuse,AltmanVoskReview,SerbynMoore,AgarwalGopalakrishnan,GopalakrishnanMueller,HuseNandkishore,SerbynPapicPRX}. 
For the random field Heisenberg chain it has been suggested, in
particular, that there is a transition at a finite disorder strength
between an ergodic phase and a non-ergodic many-body localized (MBL)
phase
\cite{OganesyanHuse,PalHuse,Luitz1,Luitz2,SerbynPapic,AndraschkoEnssSirker,EnssAndraschkoSirker,BarLevCohen}. 
Experimentally, the localization of interacting particles in quasi
one-dimensional geometries has been studied in ultracold fermionic
gases and in systems of trapped ions
\cite{SchreiberHodgman,SmithLee}. Quite recently, experimental studies have been extended to two-dimensional systems. 
In particular, the decay of an imbalance in the occupation of even and
odd sites (see Fig.~\ref{Fig1}) in fermionic chains as a function of
the interchain coupling and the onsite Hubbard interaction has been
investigated. For the case of identical disorder in the coupled chains
it has been suggested that the system remains localized in the
non-interacting Anderson case when interchain couplings are turned on
while the coupling leads to delocalization in the interacting case
\cite{SchneiderBloch2}. Theoretically, the decay rate in coupled
interacting Hubbard chains has been addressed by perturbative means
\cite{Prelovsek2016}.
For Hubbard chains, evidence for non-ergodic behavior has been found
at strong disorder in numerical simulations.\cite{MondainiRigol} A
non-ergodic phase was also found in the two-dimensional
Anderson-Hubbard model with independent disorder for each spin species
using a self-consistent perturbative approach.\cite{BarLevReichman}
For coupled chains of non-interacting spinless fermions with {\it
independent} potential disorder in each chain it has been found that
interchain coupling can both strengthen or weaken Anderson
localization, depending on the number of legs and the ratio of inter-
to intrachain coupling.\cite{WeinmannEvangelou}

The purpose of this paper is to investigate quench dynamics in
tight-binding models of fermionic chains with {\it identical}
potential disorder for different initial states and interchain
couplings, both in the non-interacting and in the interacting
case. Our study relies on analytical arguments as well as on exact
diagonalizations of finite systems. Our main results are: For the
initial state used in the experiment of
Ref.~\onlinecite{SchneiderBloch2} (see Fig.~\ref{Fig1}(a,b)) we
confirm that the dynamics in the non-interacting case is separable and
completely independent of the coupling between the chains. The
Anderson localized state is fully stable because perpendicular
interchain couplings for this particular setup are ineffective. For
generic interchain couplings and generic initial states, on the other
hand, we find that the occupation imbalance does decay both in the
Anderson and the MBL phase.

Our paper is organized as follows: In Sec.~\ref{Model} we define the
fermionic Hubbard models, initial states, and order parameters
investigated. In Sec.~\ref{Clean} we obtain analytical results for the
time dependence of the order parameters after a quench in the
non-interacting, clean limit. Based on the initial state and the
geometry of the interchain couplings we make several general
observations in Sec.~\ref{General} on whether or not the coupling
between the chains will affect the dynamics. Specific cases of
disordered free fermionic ladder models are considered in
Sec.~\ref{fF} while numerical results for interacting systems are
provided in Sec.~\ref{Int}. In addition to the order parameters, we
also consider the time evolution of the entanglement entropy of the
ladder system, see Sec.~\ref{Ent}. Finally, we summarize and conclude.

\section{Model}
\label{Model}
We consider a model of coupled fermionic Hubbard chains
\begin{eqnarray}
\label{model}
&& H=-J\sum_{i,j=1;\sigma}^{L_x-1,L_y} (c^{\dagger}_{i,j,\sigma}c_{i+1,j,\sigma}+c^{\dagger}_{i+1,j,\sigma}c_{i,j,\sigma}) \\ \nonumber
&&- J_\perp \sum_{i,j=1;\sigma}^{L_x,L_y-1} (c^{\dagger}_{i,j,\sigma}c_{i,j+1,\sigma}+c^{\dagger}_{i,j+1,\sigma}c_{i,j,\sigma}) \\
&&- J_d \sum_{i,j=1;\sigma}^{L_x-1,L_y-1}\!\!\!\!\! (c^{\dagger}_{i,j,\sigma}c_{i+1,j+1,\sigma}+c^{\dagger}_{i+1,j,\sigma}c_{i,j+1,\sigma}+h.c.) \nonumber \\
&&+ U \sum_{i,j=1}^{L_x,L_y} \left(n_{i,j,\uparrow}n_{i,j,\downarrow}-\frac{1}{2}\right) +\sum_{i,j=1;\sigma}^{L_x,L_y} D_{i} n_{i,j,\sigma} \nonumber
\end{eqnarray}
with $L_x$ sites along the $x$ direction and $L_y$ sites along the
$y$-direction. $c^{(\dagger)}_{i,j,\sigma}$ annihilates (creates) an
electron with spin $\sigma=\uparrow,\downarrow$ at site $(i,j)$, and
the local density operator is given by
$n_{i,j,\sigma}=c^{\dagger}_{i,j,\sigma}c_{i,j,\sigma}$. $J$ is the
hopping amplitude along the $x$-direction, $J_\perp$ the hopping
amplitude along $y$, and $J_d$ a diagonal hopping amplitude. $U$ is
the onsite Hubbard interaction. The random disorder potential $D_i$
only depends on the position along the $x$-direction. It is the same
for all sites with the same index $j$. We assume open boundary
conditions in both directions. In the numerical calculations we will
set $J=1$.

We are interested in the non-equilibrium dynamics of the disordered
fermionic Hubbard model \eqref{model} starting from a prepared initial
product state. Following recent experiments on cold fermionic gases we
consider, in particular, initial product states at quarter filling for
chains with $L_x$ even. We concentrate on two initial states. The
first one is given by
\begin{equation}
\label{Psi1}
|\Psi_1\rangle =\prod_{i=1}^{L_x/2}\prod_{j=1}^{L_y} c^\dagger_{2i-1,j} |0\rangle \,.
\end{equation}
In the following, we call this state the rung occupied state, see
Fig.~\ref{Fig1}(a,b). The second initial state we will consider is the
diagonally occupied state
\begin{equation}
\label{Psi2}
|\Psi_2\rangle =\prod_{i=1}^{L_x/2}\prod_{j=1}^{L_y/2} c^\dagger_{2i,2j}c^\dagger_{2i-1,2j-1} |0\rangle \,,
\end{equation}
depicted in Fig.~\ref{Fig1}(c,d). For free fermions the time evolution
of the order parameter will not depend on the spin. For interacting
fermions we consider the spin of the particles in the initial states
above as being completely random. 

For the initial state $|\Psi_1\rangle$ the order parameter is given by
\begin{equation}
\label{I1}
I_1 = \frac{2}{L_xL_y}\sum_{i,j} (-1)^{i+1} n_{ij}
\end{equation}
while
\begin{equation}
\label{I2}
I_2 = \frac{2}{L_xL_y}\sum_{i,j} (-1)^{i+j} n_{ij}
\end{equation}
is the order parameter for the initial state $|\Psi_2\rangle$. Here
$n_{ij}=\sum_\sigma n_{ij\sigma}$. Both order parameters are
normalized such that $\langle
I_1(0)\rangle=\langle\Psi_1|I_1|\Psi_1\rangle = 1$ and $\langle
I_2(0)\rangle= \langle\Psi_2|I_2|\Psi_2\rangle =1$. In the following,
we study the unitary time evolution of the order parameters $\langle
I_{1,2}(t)\rangle$ under the Hamiltonian \eqref{model} for different
sets of parameters.

\section{Free fermions in the clean limit}
\label{Clean}
We start with the clean free fermion case $U=0$ and $D_{i}=0$. The
Hamiltonian can then be diagonalized by Fourier transform and the time
evolution of $\langle I_{1,2}(t)\rangle$ can be calculated
analytically. The Fourier representation of the annihilation operator
for open boundary conditions is given by
\begin{equation}
\label{Fourier}
c_{ij\sigma}= \frac{2}{\sqrt{(L_x+1)(L_y+1)}}\sum_{k_x,k_y}\sin k_x \sin k_y c_{k_x,k_y,\sigma}. 
\end{equation}
The wave vectors are quantized according $k_x=n\pi/(L_x+1)$ and
$k_y=m\pi/(L_y+1)$ with $n=1,\cdots,L_x$; $m=1,\cdots,L_y$. Unitary
time evolution results in
\begin{equation}
\label{ckt}
c_{k_x,k_y,\sigma}(t)=\exp(-\im\varepsilon_{k_x,k_y,\sigma}t)c_{k_x,k_y,\sigma}
\end{equation}
where the dispersion for model \eqref{model} reads
\begin{equation}
\label{disp}
\varepsilon_{k_x,k_y}= 2J\cos k_x +2J_\perp\cos k_y + 4J_d \cos k_x \cos k_y
\end{equation}
and is independent of the spin index $\sigma$. Using the Fourier
expansion \eqref{Fourier} for the order parameter \eqref{I1} we find
\begin{eqnarray}
\label{I1t}
 \langle I_1(t)\rangle &=& \frac{1}{L_xL_y}\sum_{n,m=1}^{L_x,L_y} \exp[\im(\varepsilon_{n,m}-\varepsilon_{L_x+1-n,m})] \\ 
&=& \frac{1}{L_xL_y}\sum_{k_x,k_y} \exp\left[4\im t\cos k_x \left(J+2J_d\cos k_y\right)\right] \nonumber \\
&\to& \frac{1}{\pi}\int_0^\pi dk_x\exp[4\im tJ\cos k_x]J_0(8J_dt\cos k_x) \nonumber
\end{eqnarray}
where we have taken the thermodynamic limit, $L_x,L_y\to\infty$, in
the last line with $J_0$ being the Bessel function of the first
kind. Without the diagonal couplings ($J_d=0$) as in the experiment of
Ref.~\onlinecite{SchneiderBloch2} we find, in particular,
\begin{equation}
\label{I1t2}
\langle I_1(t)\rangle=J_0(4Jt)\sim (2\pi Jt)^{-1/2}
\end{equation}
in the thermodynamic limit while
$\langle I_1(t)\rangle\sim (JJ_d)^{-1/2}/t$ for $J_d\neq 0$.

Importantly, the result for the the initial state $|\Psi_1\rangle$ is
always independent of the coupling in the transverse direction
$J_\perp$. Without diagonal couplings we have a fine-tuned setup where
$\langle I_1(t)\rangle$ is {\it identical} to the result for a single
chain. While a generic coupling between the chains will typically lead
to a faster dephasing and therefore to a faster decay of the order
parameter this is not the case in such a fine-tuned setup. 

For a finite number of legs one can also prevent the order parameter
$\langle I_1(t)\rangle$ from decaying completely by fine-tuning the
diagonal coupling $J_d$. This happens if for any of the allowed wave
vectors $k^{(m)}_y = m\pi/(L_y+1)$ the diagonal coupling is chosen such
that $J_d=-J/(2\cos k^{(m)}_y)$. For an infinite two-leg ladder, for
example, we find $\lim_{t\to\infty}\langle I_1(t)\rangle=1/2$ if
$J_d=\pm J$ because $\cos k^{(m)}_y=\pm 1/2$ in this case.

The behavior of the order parameter \eqref{I2} for the diagonal initial state
\eqref{Psi2}, on the other hand, is very different.  In this case we find 
\begin{eqnarray}
\label{I2t}
&&\langle I_2(t)\rangle = \frac{1}{L_xL_y}\sum_{k_x,k_y} \exp\left[4it\left(J\cos k_x +J_\perp\cos k_y\right)\right] \nonumber \\
&&\stackrel{L_x,L_y\to\infty}{\to} J_0(4Jt)J_0(4J_\perp t)\sim (JJ_\perp)^{-1/2}(2\pi t)^{-1} 
\end{eqnarray}
even without diagonal couplings. For the infinite two-dimensional
lattice ($J_\perp\neq 0$, $J_d=0$) the order parameter is decaying $\sim 1/t$
for the diagonal initial state $|\Psi_2\rangle$ as compared to the
$1/\sqrt{t}$ decay for the initial state $|\Psi_1\rangle$. The
diagonally occupied state is thus a more generic initial state where a
crossover from one- to two-dimensional behavior for free fermions does
occur if the chains are coupled by a perpendicular hopping term.

\section{General results for fermionic chains with identical disorder}
\label{General}
In this section we want to provide some general arguments to show why
the time evolution of the order parameter $\langle I_1(t)\rangle$ of
the ladder system can be one-dimensional even in the presence of
interchain couplings {\it and} disorder. We concentrate here on the
system without the diagonal hopping terms ($J_d=0$) which will always
make the system two-dimensional and which are not part of the
experimental setup in Ref.~\onlinecite{SchneiderBloch2}.

First, we perform a Fourier transform along the direction of the
interchain couplings $J_\perp$. Note that all sites for a given index
$i$ along the $x$-direction have the same potential. The Hamiltonian
\eqref{model} can then be written as $H=H_J + H_{J_\perp} + H_D = \sum_i
h_i$ with
\begin{eqnarray}
\label{partialFT}
h_i &=& J\sum_{k_y} (c^\dagger_{i,k_y}c_{i+1,k_y}+h.c.) \\
&+& 2J_\perp\sum_{k_y} \cos k_y n_{i,k_y} +D_i\sum_{k_y}n_{i,k_y}. \nonumber
\end{eqnarray}
Similarly, we can write the order parameter as
\begin{equation}
I_1 = \frac{2}{L_xL_y}\sum_{i,k_y} (-1)^{i+1} n_{i,k_y}.
\end{equation}
In this representation it is immediately clear that
$[H_{J_\perp},H_J]=[H_{J_\perp},H_D]=[H_{J_\perp},n_{i,k_y}]=0$ thus
\begin{equation}
c^\dagger_{i,k_y}c_{i,k_y}(t)=\e^{-i(H_J+H_D)t}c^\dagger_{i,k_y}c_{i,k_y}\e^{i(H_J+H_D)t}.
\end{equation}
For free fermions $\langle I_1(t)\rangle$ is therefore independent of
$J_\perp$ even in the presence of disorder. This order parameter will
therefore {\it always appear to indicate} that the Anderson localized phase
is stable against perpendicular interchain couplings.

If, on the other hand, diagonal hoppings are included then $H_{J_d}$
does not commute with $n_{i,k_y}$. In this generic situation the
Anderson localized chain will be affected by the diagonal interchain
couplings $J_d$. We analyze several examples in more detail in the
next section. Similarly, introducing a Hubbard interaction $U$ implies
that $H_{J_\perp}$ does not commute with the rest of the Hamiltonian
anymore. On this level, the roles played by $H_{J_d}$ and $H_U$ are
similar: both break the fine-tuned symmetry which make the disordered
system behave completely one-dimensional even in the presence of
couplings $J_\perp$ between the chains. Without the diagonal hopping
terms the initial state $|\Psi_1\rangle$ together with the order
parameter $\langle I_1(t)\rangle$ are thus not suitable to study the
{\it generic differences} between Anderson and many-body localization in
coupled chains with identical disorder.

\begin{figure}
  \includegraphics*[width=0.99\columnwidth]{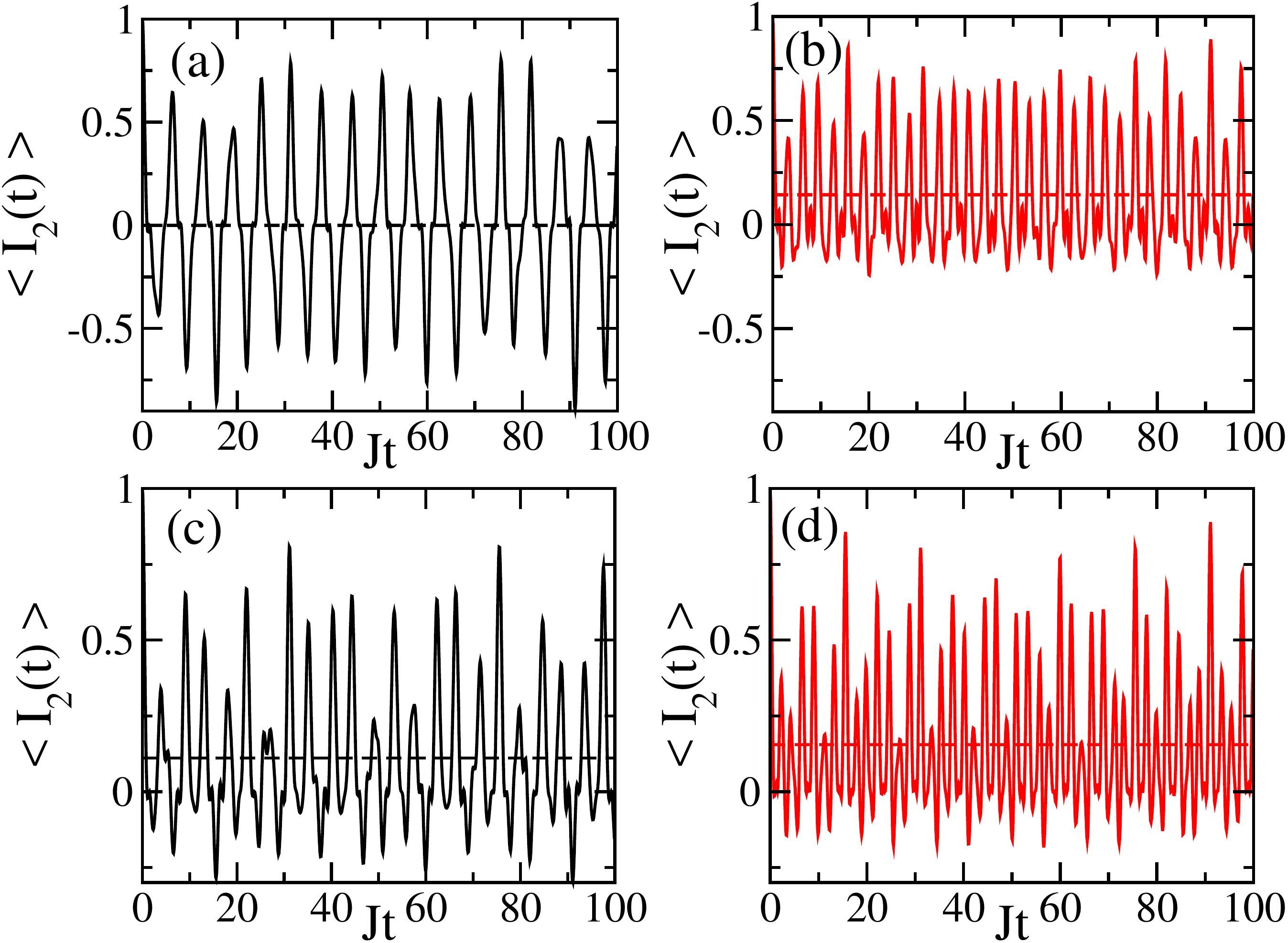}
  \caption{(Color online) $\langle I_2(t)\rangle$ for infinite binary
  disorder. Results for two-leg ladders (a,b) and three-leg ladders
  (c,d) with $200$ sites in the chain direction are
  presented. Interchain couplings are (a,c) $J_\perp =0.5$ and (b,d)
  $J_\perp=1$. Averages over $2000$ samples are shown. The dashed
  lines denote the long-time averages, see text.}
\label{Fig_bin_dis}
\end{figure}
\section{Free fermions with binary and box disorder}
\label{fF}
In this section we want to consider specific examples for the 
Hamiltonian \eqref{model} with $U=0$ and different types of disorder.

\subsection{Free fermions with infinite binary disorder}
Apart from the clean non-interacting case we can also study the case
of binary disorder, $D_i=\pm D$, in the limit $D\to\infty$
analytically. We consider, in particular, a ladder with $L_y$ legs and
$J_d=0$ in the limit $L_x\to\infty$. The infinite binary potential
along the $x$-direction then splits the ladder system into decoupled
finite clusters with equal potential. The disorder averaged time
evolution of the system is then given by a sum of the time evolution
of open clusters $I_{1,2}^\ell(t)$ with length $\ell$ along the
$x$-direction and width $L_y$ weighted by their probability of
occurence $p_\ell=\ell/2^{\ell+1}$ with $\sum_\ell p_\ell =1$
\cite{AndraschkoEnssSirker}. For infinite binary disorder the disorder average 
of the order parameters is therefore given by
\begin{equation}
\label{inf_dis}
\langle I_{1,2}^{D=\infty}(t)\rangle =\sum_{\ell=1}^\infty p_\ell \langle I_{1,2}^\ell (t)\rangle.
\end{equation}

For the rung occupied state, $\langle I_1^\ell(t)\rangle$ does not depend on
$J_\perp$. The result for $I_{1}^{D=\infty}(t)$ is therefore exactly
the same as for a single chain. In particular, only clusters with
$\ell$ odd give a contribution $\overline{I_1^{\ell\;
\textrm{odd}}} =1/\ell$ to the time average so that 
$\overline{I_{1}^{D=\infty}} =\sum_{\ell\;\textrm{odd}}
\frac{p_\ell}{\ell}=1/3$. There is no dephasing in the case of
infinite binary disorder. $\langle I_{1}^{D=\infty}(t)\rangle$ does show persistent
oscillations around the time average $\overline{I_{1}^{D=\infty}}=1/3$
\cite{AndraschkoEnssSirker,EnssAndraschkoSirker}.

For the diagonally occupied state the situation is very different. Let
us first consider the case of an even number of legs, i.e., $L_y$
even. In this case every decoupled cluster with equal potential of
size $\ell\times L_y$ will have $\ell L_y/2$ fermions. For the generic
case $J\neq J_\perp$ the order parameter $I_2^\ell(t)$ will then show
persistent oscillations around zero for all cluster lengths $\ell$
resulting in $\overline{I_2^{D=\infty}}=0$. For $J=J_\perp$, on the
other hand, clusters with length $\ell=n(L_y+1)-1$; $n=1,2,\cdots$
will give a contribution $\overline{I_2^{\ell}} =1/\ell$ to the time
average so that $\overline{I_{2}^{D=\infty}} =\sum_{\ell}
\frac{p_\ell}{\ell}=\sum_{n=1}^\infty 2^{-n(L_y+1)}=1/(2^{L_y+1}-1)$. 
For $L_y$ odd and $J\neq J_\perp$ all clusters with odd length $\ell$
will give a contribution $1/(L_y\ell)$ so that
$\overline{I_{2}^{D=\infty}} =1/(3L_y)$. For $J=J_\perp$ and $L_y$ odd,
clusters of length $\ell=n(L_y+1)-1$, $n=1,2,\cdots$ will give a
$1/\ell$ contribution to the time average while all other odd clusters
will contribute $1/(L_y\ell)$ giving rise to a time average 
\begin{equation}
\label{I2odd}
\overline{I_2^{D=\infty}}=\frac{1}{3L_y}+\frac{1-1/L_y}{2^{L_y+1}-1}.
\end{equation}
In Fig.~\ref{Fig_bin_dis} these analytically obtained long-time
averages are compared to numerical data. For the two-leg ladder with
$J_\perp =0.5J$, see Fig.~\ref{Fig_bin_dis}(a), the long-time average
is zero while for $J_\perp = J$, see Fig.~\ref{Fig_bin_dis}(b), we
have $\overline{I_2}=1/7$. For the three-leg ladder we find, on the
other hand, $\overline{I_2}=1/9$ and $\overline{I_2}=7/45$,
respectively.

To summarize, there is an interesting even/odd effect for the
diagonally occupied state with $\overline{I_2^{D=\infty}}=0$ for $L_y$
even and $\overline{I_2^{D=\infty}}=\frac{1}{3L_y}$ for a generic
interchain coupling $J_\perp\neq J$. In the following subsection we
will see that these even/odd effects do persist for finite box
disorder.

\begin{figure}
  \includegraphics*[width=0.99\columnwidth]{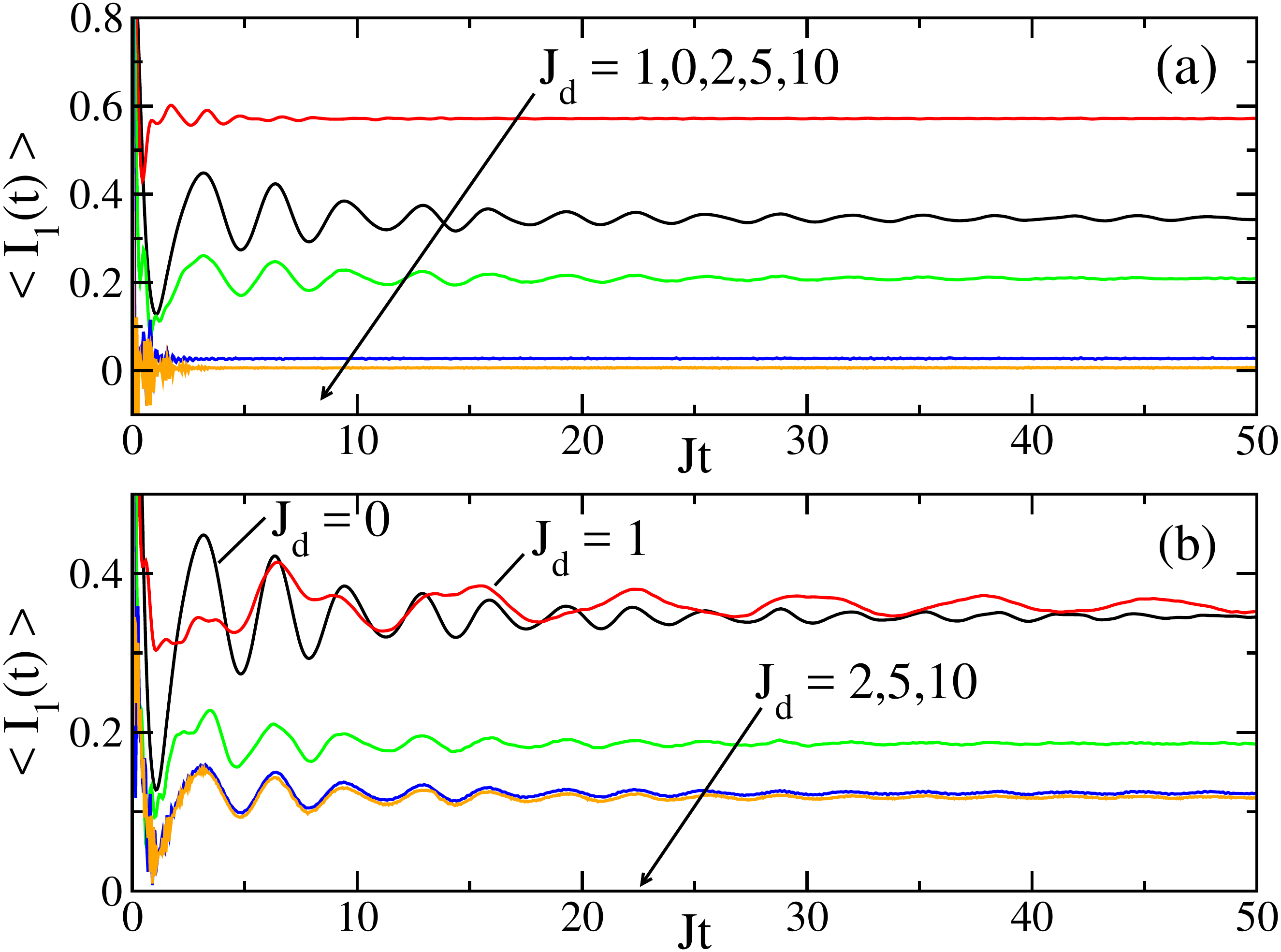}
  \caption{(Color online) $\langle I_1(t)\rangle$ for (a) two-leg
  ladders and (b) three-leg ladders with $200$ sites along the chain
  direction, box disorder $D=5$ and different diagonal couplings $J_d$
  as indicated. The results are independent of the perpendicular
  interchain coupling $J_\perp$. Averages over $1000$ samples are
  shown with statistical errors of the order of the line width.}
\label{Fig2}
\end{figure}
\begin{figure}
  \includegraphics*[width=0.99\columnwidth]{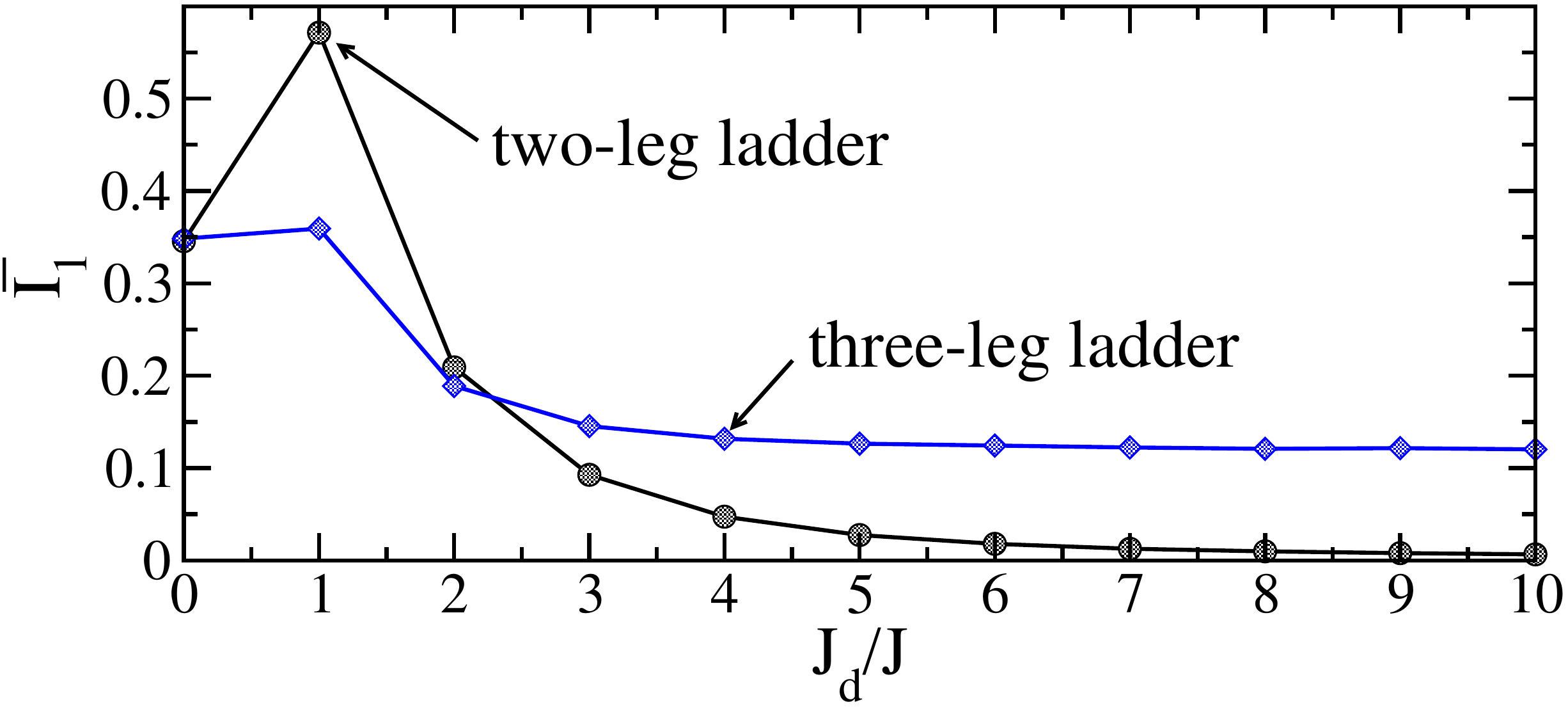}
  \caption{(Color online) Long-time averages $\overline{I_1}$ for the
  data shown in Fig.~\ref{Fig2}. For the two-leg ladder the data are
  consistent with a power-law decay to zero for $J_d\to\infty$ while a
  power-law fit yields $\overline{I_1}(J_d\to\infty)\approx 0.119$ for
  the three-leg ladder.}
\label{Fig4}
\end{figure}
\subsection{Free fermions with box disorder}
Here we want to present numerical results for non-interacting ladders
with disorder drawn from a box distribution $D_i\in [-D,D]$. Because
the system is non-interacting, calculating the order parameters
$\langle I_{1,2}(t)\rangle$ reduces to an effective one-particle
problem which can be solved numerically for large system sizes. We
start from the initial $L_xL_y/2$ one-particle states in position
representation and time evolve each of these states using the
Hamiltonian
\eqref{model} with $U=0$. The order parameters are then simply given 
by the sum of the order parameters for each one particle wave
function. We have checked that the numerical data agree with the
analytical solutions in Sec.~\ref{Clean} for the clean case and that
$\langle I_1(t)\rangle$ is indeed independent of $J_\perp$ for all
disorder strengths.

We start by presenting data in Fig.~\ref{Fig2}(a) for the time
evolution in two-leg ladders prepared in the rung occupied initial
state. As discussed in section Sec.~\ref{General} the results are
independent of $J_\perp$. While the order parameter is increased for
$J_d=J$, stronger diagonal couplings lead to a decrease and the data
are consistent with $\overline{I_1}\to 0$ for $J_d\to\infty$, see
Fig.~\ref{Fig4}. In Fig.~\ref{Fig2}(b) data for the same parameters
but for three-leg ladders are shown. The results are quite different
from the two-leg case. While the results are again independent of
$J_\perp$, we now find that the long-time average $\overline{I_1}$
remains non-zero even for strong interchain couplings $J_d/J\gg 1$,
see Fig.~\ref{Fig4}. The long-time behavior is thus quite different
for ladders with an even or an odd number of legs. Similar to the case
of infinite binary disorder we expect that for ladders with an odd
number of legs the long-time average $\overline{I_1}$ decreases with
the number of legs. Coupling an infinite number of Anderson localized
chains with identical disorder in a generic way will thus lead to a
complete destruction of the order parameter.

\begin{figure}
  \includegraphics*[width=0.99\columnwidth]{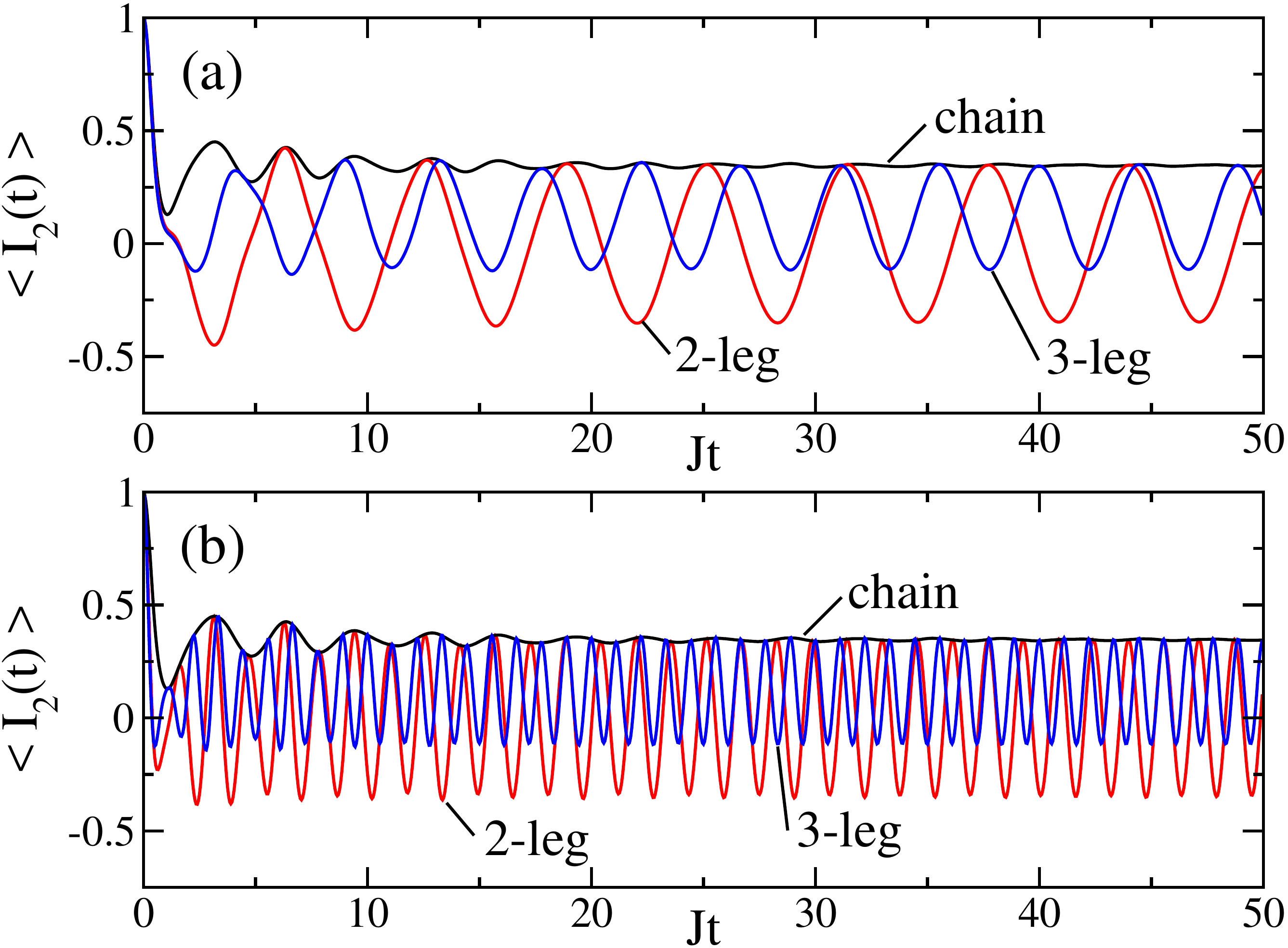}
  \caption{(Color online) $\langle I_2(t) \rangle$ for a chain, a two-leg, and a
  three-leg ladder with $200$ sites along the chain direction and
  $D=5$. (a) $J_\perp =0.5$, and (b) $J_\perp =2$. Averages over
  $1000$ samples are shown. $\overline{I_2}\neq 0$ for the single
  chain and the three-leg ladder while $\overline{I_2}=0$ for the
  two-leg ladder.}
\label{Fig5}
\end{figure}

Next, we present data for the diagonally occupied initial state in
Fig.~\ref{Fig5}.
The results are qualitatively similar to the case of infinite binary
disorder solved analytically in the previous section. In particular,
we find that for generic interchain couplings $J_\perp$ the long-time
average $\overline{I_2}$ is zero for an even number of legs while it
is non-zero for an odd number of legs.

\section{Interacting ladder models}
\label{Int}
We now turn to a numerical study of the interacting case. Here we are
limited to the exact diagonalization of rather small two-leg
ladders. While the system sizes could, in principle, be increased the
substantial number of samples required to obtain disorder averages
with small statistical errors is a further limiting factor in
practice. Nevertheless, even these small systems show behavior which
is qualitatively consistent with the experimental results in
Ref.~\onlinecite{SchneiderBloch2}.
\subsection{Spinful fermions}
We concentrate first on spinful fermions on a two-leg ladder with
onsite Hubbard interaction $U$. For a $4\times 2$ ladder with
$n_\uparrow=n_\downarrow=2$ the Hilbert space has dimension
${\binom{8}{2}}^2=784$. We find that in the interacting case a much
larger number of samples than in the non-interacting case is required
(by at least a factor of $10$) to obtain the same accuracy for the
disorder average. For a $4\times 2$ ladder this is still easily
achievable while already for a $6\times 2$ ladder with
$n_\uparrow=n_\downarrow=3$ the Hilbert space dimension is
${\binom{12}{3}}^2=48400$, and an enormous amount of computing
resources would be required. Instead, we will also present results for
a $6\times 2$ ladder with $n_\uparrow=4$ and $n_\downarrow=2$ with
Hilbert space dimension ${\binom{12}{2}}{\binom{12}{4}}=32670$.

\begin{figure}
  \includegraphics*[width=0.99\columnwidth]{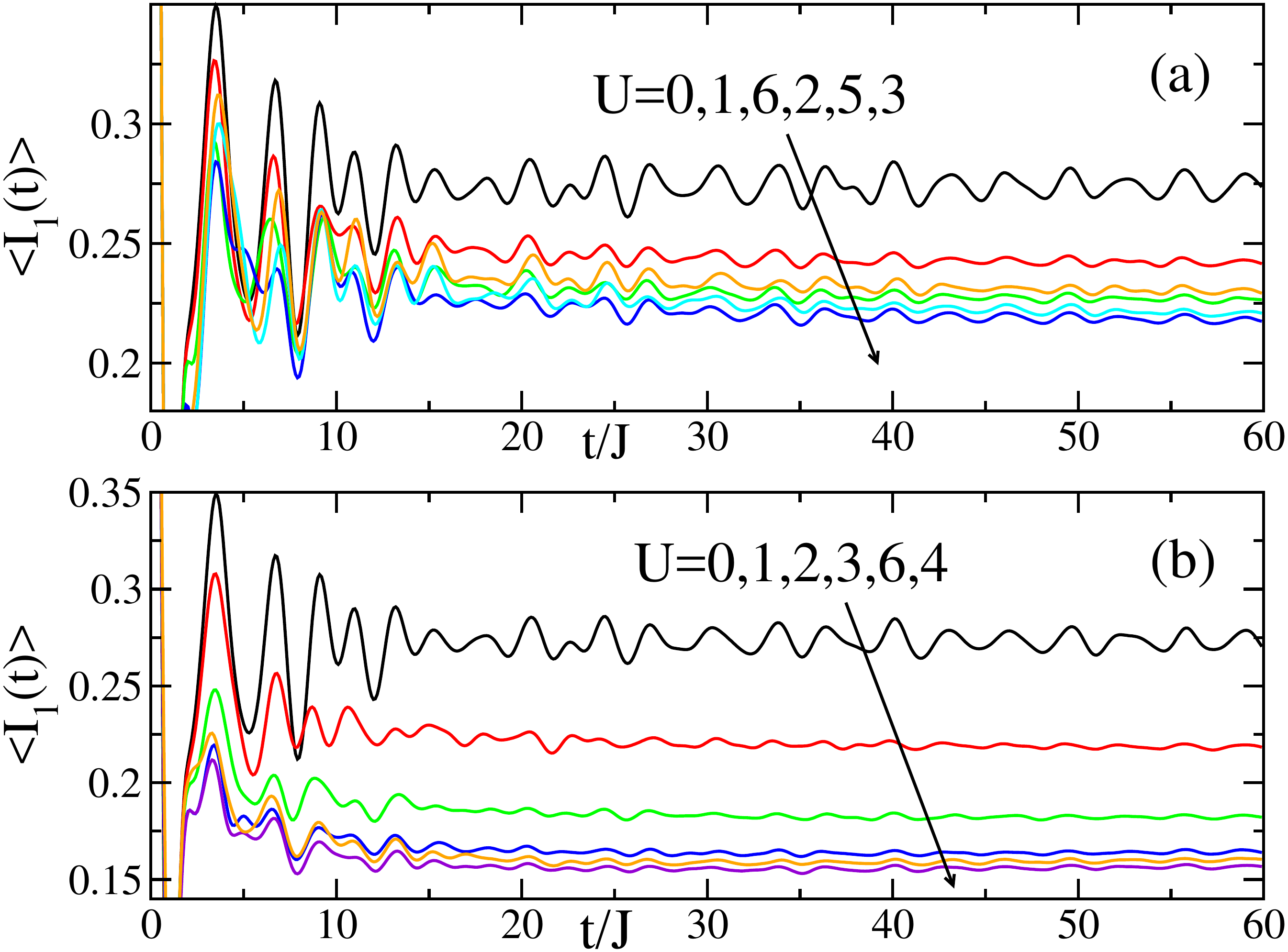}
  \caption{(Color online) $\langle I_1(t)\rangle$ for a $4\times 2$
  ladder of spinful fermions with disorder $D=2.5$. Averages over
  $200\, 000$ samples are shown. (a) $J_\perp =0.1$, and (b) $J_\perp
  =1$. The Hubbard interactions $U$ are indicated.}
\label{Fig6}
\end{figure}

\begin{figure}
  \includegraphics*[width=0.99\columnwidth]{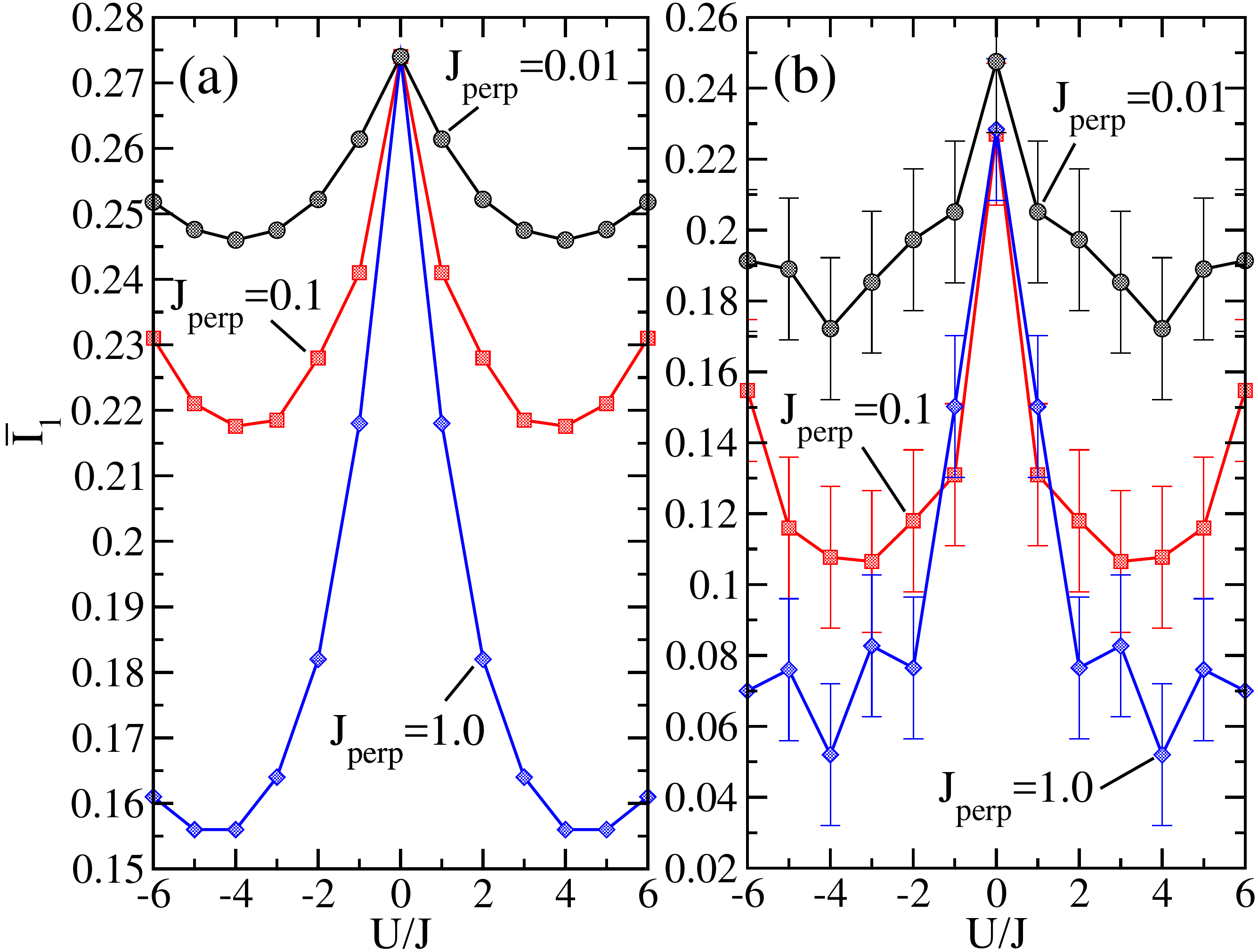}
  \caption{(Color online) Long-time average $\overline{I_1}$ at
  disorder $D=2.5$ for (a) a $4\times 2$ ladder with
  $n_\uparrow=n_\downarrow =2$ for $200\,000$ samples, see
  Fig.~\ref{Fig6}, and (b) a $6\times 2$ ladder with $n_\uparrow=4$
  and $n_\downarrow=2$ for $400$ samples. Note that the dependence on
  $|U|/J$ is non-monotonic.}
\label{Fig7}
\end{figure}

In Fig.~\ref{Fig6} the order parameter $\langle I_1(t)\rangle$ for the
$4\times 2$ ladder is shown for different interaction strengths $U/J$.
Both for weak and for strong interchain coupling, increasing the
Hubbard interaction initially leads to a decrease of the long time
average $\overline{I_1}$ with a minimum at $|U|/J\sim 4-5$, see
Fig.~\ref{Fig7}(a).
For even larger interaction strengths the long-time average increases
again leading to a characteristic shape of the imbalance versus
$|U|/J$ curve qualitatively consistent with the experimental data
obtained in Ref.~\onlinecite{SchneiderBloch2}. The same is true for
the $6\times 2$ ladder, see Fig.~\ref{Fig7}(b), although the small
number of samples we have simulated leads to relatively large error
bars. Note that the arguments presented in
Ref.~\onlinecite{EnssSirker} for the $U\to -U$ symmetry in such
quenches for clean Hubbard models remain valid even if potential
disorder is included. The sign of $U$ does not affect the quench
dynamics.

\begin{figure}
  \includegraphics*[width=0.99\columnwidth]{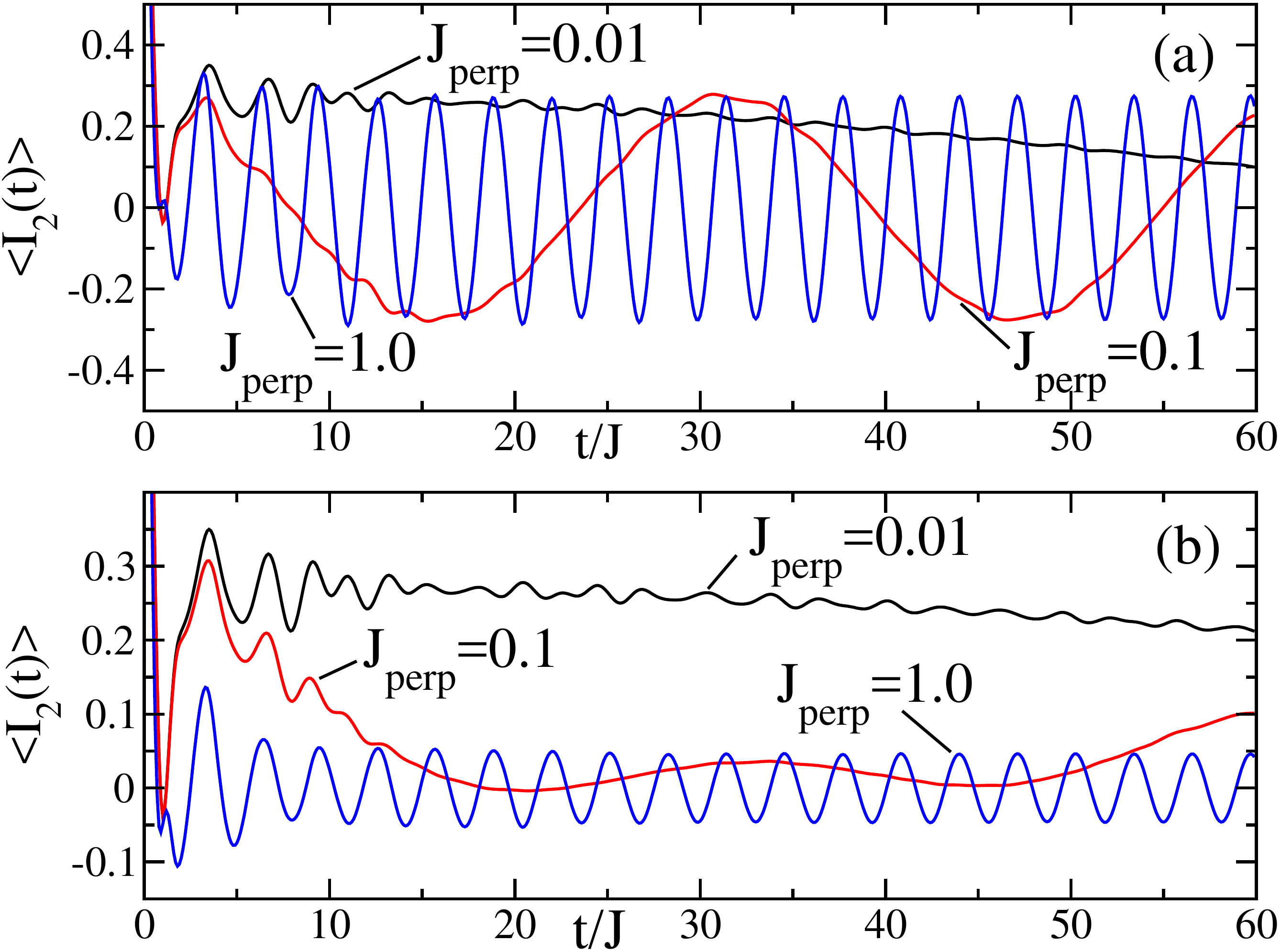}
  \caption{(Color online) $\langle I_2(t)\rangle$ for a $4\times 2$
  ladder of spinful fermions with disorder $D=2.5$. Averages over
  $500\, 000$ samples are shown for (a) $U=0$, and (b) $U=4$. The
  interchain coupling $J_\perp$ is indicated on the graph. $\langle
  I_2(t)\rangle$ for $J_\perp=0.1$ in (b) also starts to oscillate around
  zero for longer times (data not shown).}
\label{Fig7_diag}
\end{figure}

Results for the diagonal initial state $|\Psi_2\rangle$ are shown in
Fig.~\ref{Fig7_diag}.
For $U=0$, see Fig.~\ref{Fig7_diag}(a), we obtain results for the
$4\times 2$ ladder which show qualitatively the same behavior as the
ones already presented in Fig.~\ref{Fig5} for much larger
ladders. $\langle I_2(t)\rangle$ for $J_\perp\neq 0$ oscillates around
zero with $J_\perp$ determining the oscillation frequency. While the
oscillation amplitude around zero is modified for $U=4$, see
Fig.~\ref{Fig7_diag}(b), there is otherwise no qualitative difference
between the non-interacting and the interacting case. For a given
coupling strength $J_\perp$ the time scale for the initial decay of
$\langle I_2(t)\rangle$ is of the same order. For interchain coupling
$J_\perp =1$ we observe, in particular, an almost complete decay of
the order parameter on a time scale of order $J$ in both cases.

\subsection{Spinless fermions}
\begin{figure}
  \includegraphics*[width=0.99\columnwidth]{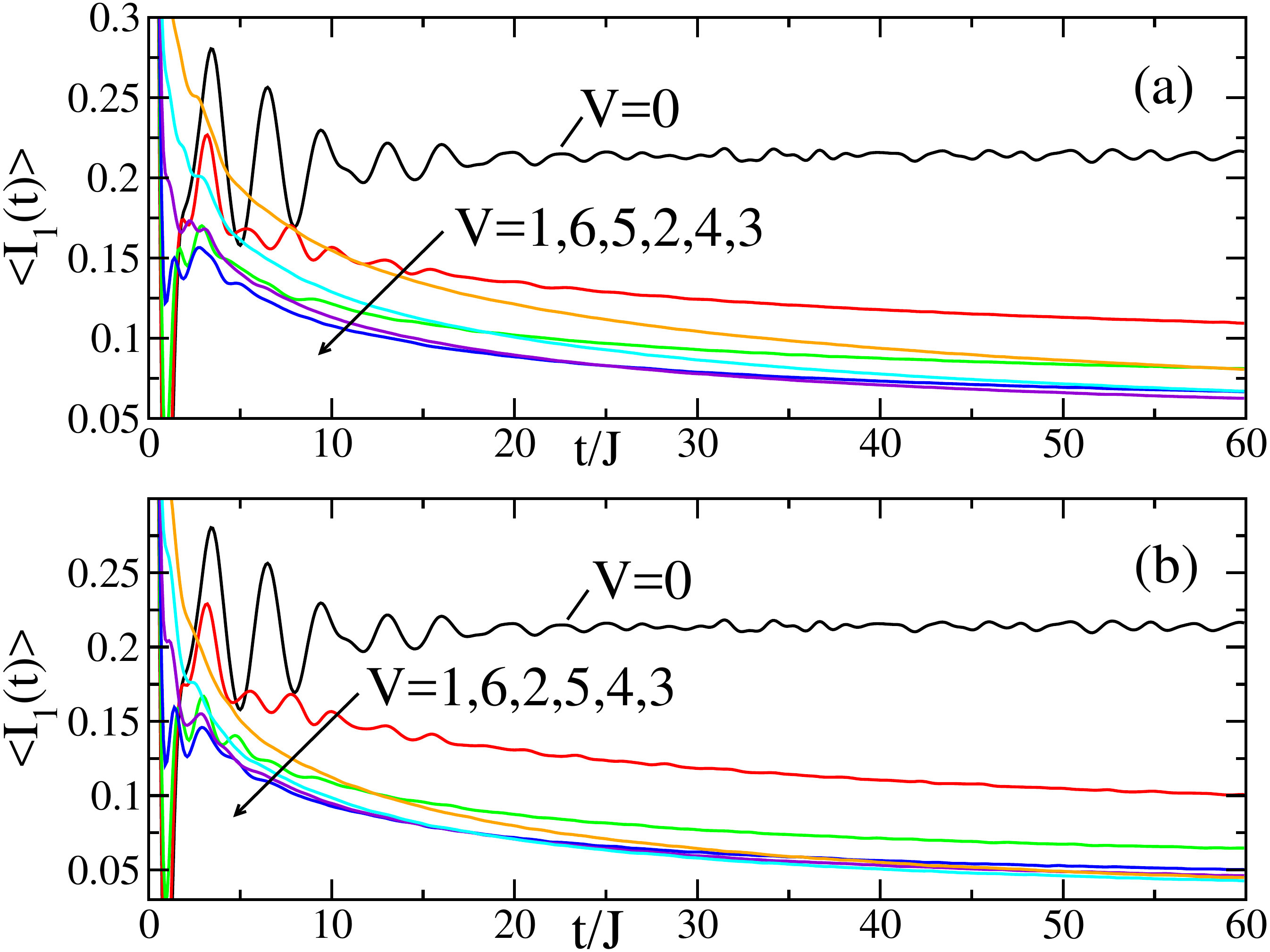}
  \caption{(Color online) $\langle I_1(t)\rangle$ for a $8\times 2$
  ladder of spinless fermions with disorder $D=2.5$. $2000$ to $20\,000$
  samples are used. (a) $J_\perp =0.1$, and (b) $J_\perp =1$. The
  nearest-neighbor interactions $V$ are indicated.}
\label{Fig8}
\end{figure}

While our numerical results for spinful $4\times 2$ and $6\times 2$
ladders demonstrate behavior which is qualitatively consistent with
the experimental data, the system sizes are quite small. To corroborate
these results we thus also consider the case of spinless fermionic
two-leg ladders where larger system sizes can be simulated. Instead of
an onsite Hubbard interaction $U$ we now introduce a nearest-neighbor
interaction
\begin{equation}
\label{NN}
H_V = V\sum_i \left(n_{i,1}n_{i,2}+n_{i,1}n_{i+1,1}+n_{i,2}n_{i+1,2}\right).
\end{equation}
Results for a $8\times 2$ ladder with $8$ fermions are shown in
Fig.~\ref{Fig8}. As in the spinful case, the dynamics for $V=0$ is
one-dimensional and completely independent of the strength of the
interchain coupling $J_\perp$: the results for $V=0$ in
Fig.~\ref{Fig8}(a) and Fig.~\ref{Fig8}(b) are identical.
Adding nearest-neighbor interactions leads to a strong reduction of
the order parameter both for weak and strong hopping between the
chains.
\begin{figure}
  \includegraphics*[width=0.99\columnwidth]{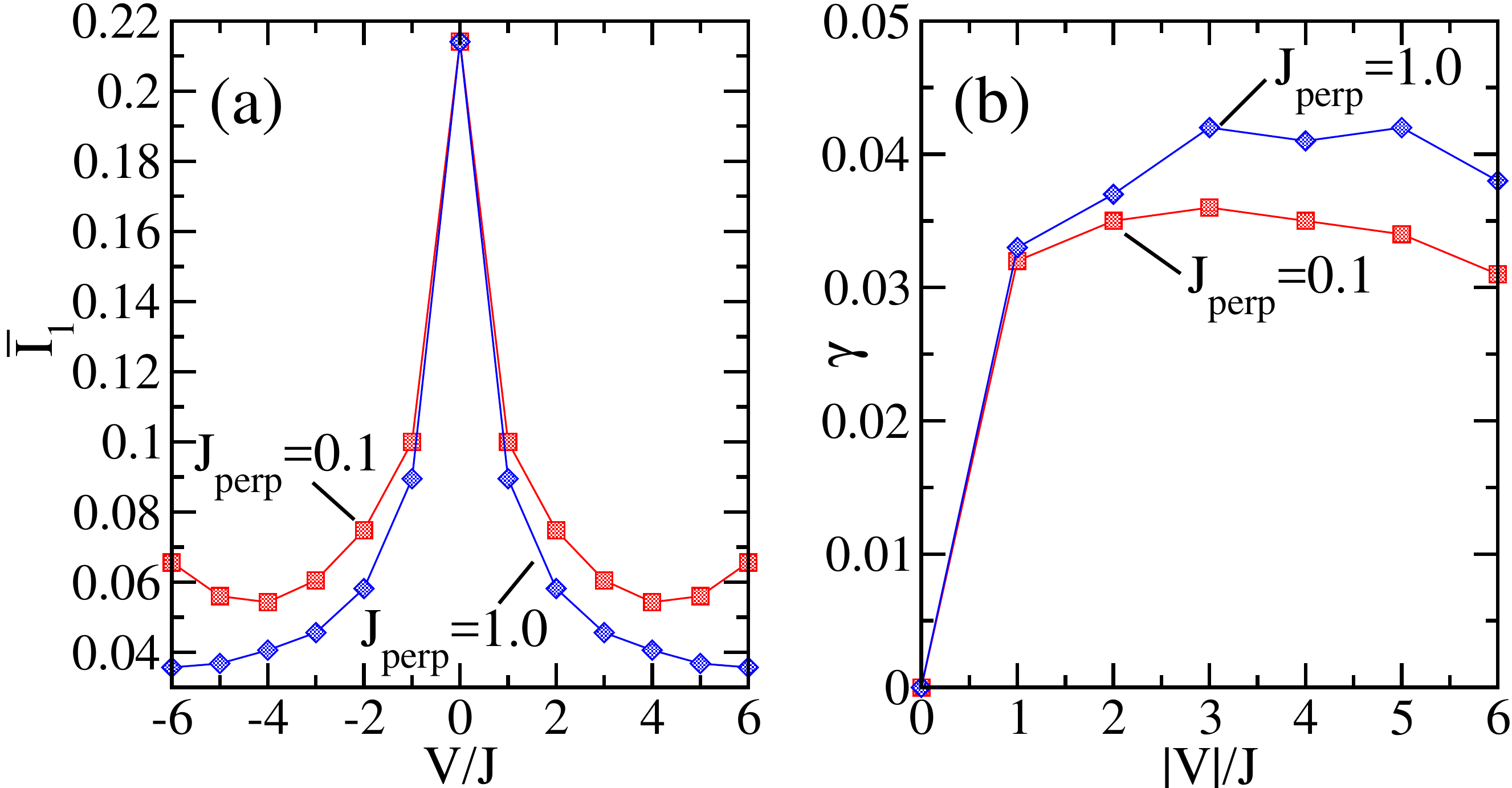}
  \caption{(Color online) Fitting the data in Fig.~\ref{Fig8}
  according to $\langle I_1(t)\rangle = \overline{I_1}+a\exp(-\gamma
  t)$ for $t>30$ allows to extract estimates for the long-time average
  $\overline{I_1}$ shown in panel (a) and the decay rate $\gamma$
  shown in panel (b).}
\label{Fig9}
\end{figure}
The decay of the order parameter at long times in the interacting case
seems to be well described by an exponential. The long-time average
and the decay rate extracted from exponential fits are shown in
Fig.~\ref{Fig9}. The results show that both the long-time average
$\overline{I_1}$ and the decay rate $\gamma$ do depend on the strength
of $J_\perp$ albeit rather weakly. For weak interchain coupling
$J_\perp=0.1$ we observe a non-monotonic dependence of
$\overline{I_1}$ on the interaction strength similar to the spinful
case. 

\section{Entanglement entropy}
\label{Ent}
In this final section we want to briefly discuss the entanglement
properties of fermionic ladders. We consider ladders where the chains
contain an even number of sites and cut the ladder into two equal
halfs, $A$ and $B$, perpendicular to the chain direction. The von
Neumann entanglement entropy is then defined as
\begin{equation}
\label{Sent}
S_{\textrm{ent}}(t) = -\tr\,\rho_A(t)\ln\rho_A(t)
\end{equation}
where $\rho_A(t)=\tr_B |\Psi_i(t)\rangle\langle\Psi_i(t)|$ is the
reduced density matrix of segment $A$. 

If we start from one of the product states $|\Psi_{1,2}\rangle$ then
the entanglement entropy for a clean ladder grows linearly in time
before saturating at a constant for times $t>L_x/(2v)$ where $L_x/2$
is the length of the segment and $v\sim 2J$ the velocity of
excitations.\cite{CalabreseCardy09} The entanglement entropy per
chain, $S_{\textrm{ent}}(t)/L_y$, in the clean case is independent of
the number of legs $L_y$ and independent of the coupling $J_\perp$
between the ladders for $J_d=0$. For spinless fermions we find, in
particular, that $S_{\textrm{ent}}(t)/L_y\sim 0.88 t$ for $t<L_x/(2v)$
consistent with the results for a single
chain.\cite{ZhaoAndraschkoSirker} Similar to the order parameter
$\langle I_1(t)\rangle$ the entanglement entropy $S_{\textrm{ent}}(t)$
for the rung occupied initial state remains independent of the
interchain coupling $J_\perp$ in the non-interacting case even if we
include disorder. Without interactions or diagonal couplings,
$S_{\textrm{ent}}(t)$ of a ladder prepared in the rung occupied
initial state is simply $L_y$ times the entanglement entropy of a
single chain. This demonstrates further that the stability of the
Anderson localized state cannot be investigated in this setup. 

\begin{figure}
  \includegraphics*[width=0.99\columnwidth]{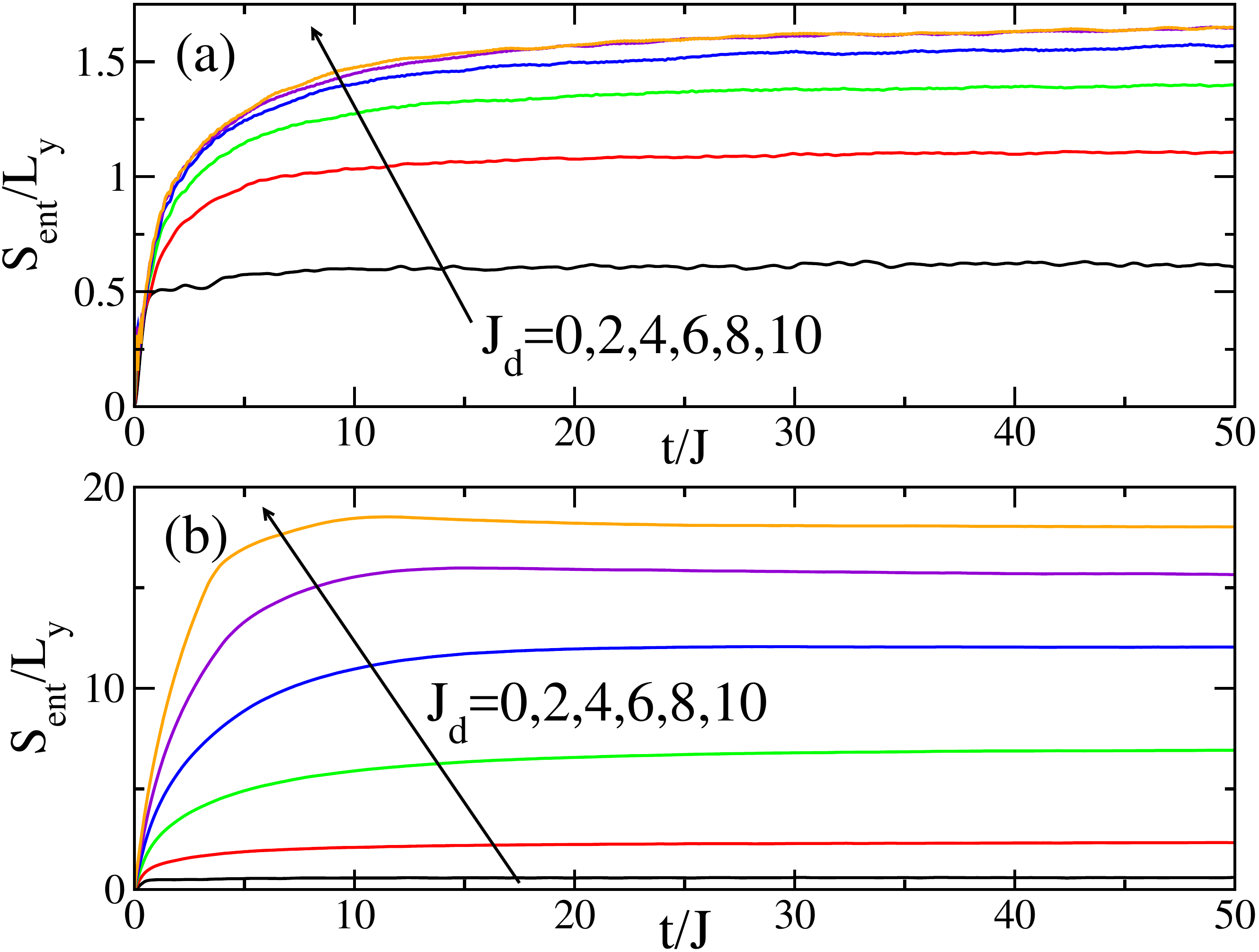}
  \caption{(Color online) Entanglement entropy per chain of free
  spinless fermions starting from the rung initial state with disorder
  $D=2.5$ and different $J_d$ for (a) a two-leg ladder, and (b) a
  three-leg ladder. The results are independent of $J_\perp$. The
  length of both ladders is $200$ sites along the chain direction;
  averages over $2000$ samples are shown.}
\label{Fig10}
\end{figure}

One way to allow for dynamics which involves the full ladder is to
include diagonal couplings. As demonstrated in Fig.~\ref{Fig10},
$S_{\textrm{ent}}(t)$ is then no longer simply given by $L_y$ times
the entanglement entropy of a single chain but rather grows more
rapidly with the number of legs as expected when moving towards a
two-dimensional system.
The entanglement entropy at long times increases monotonically with
$J_d$ up to a maximum value. The maximal value is determined by the
smaller of the two relevant length scales: the localization length and
the block size.

\begin{figure}
  \includegraphics*[width=0.99\columnwidth]{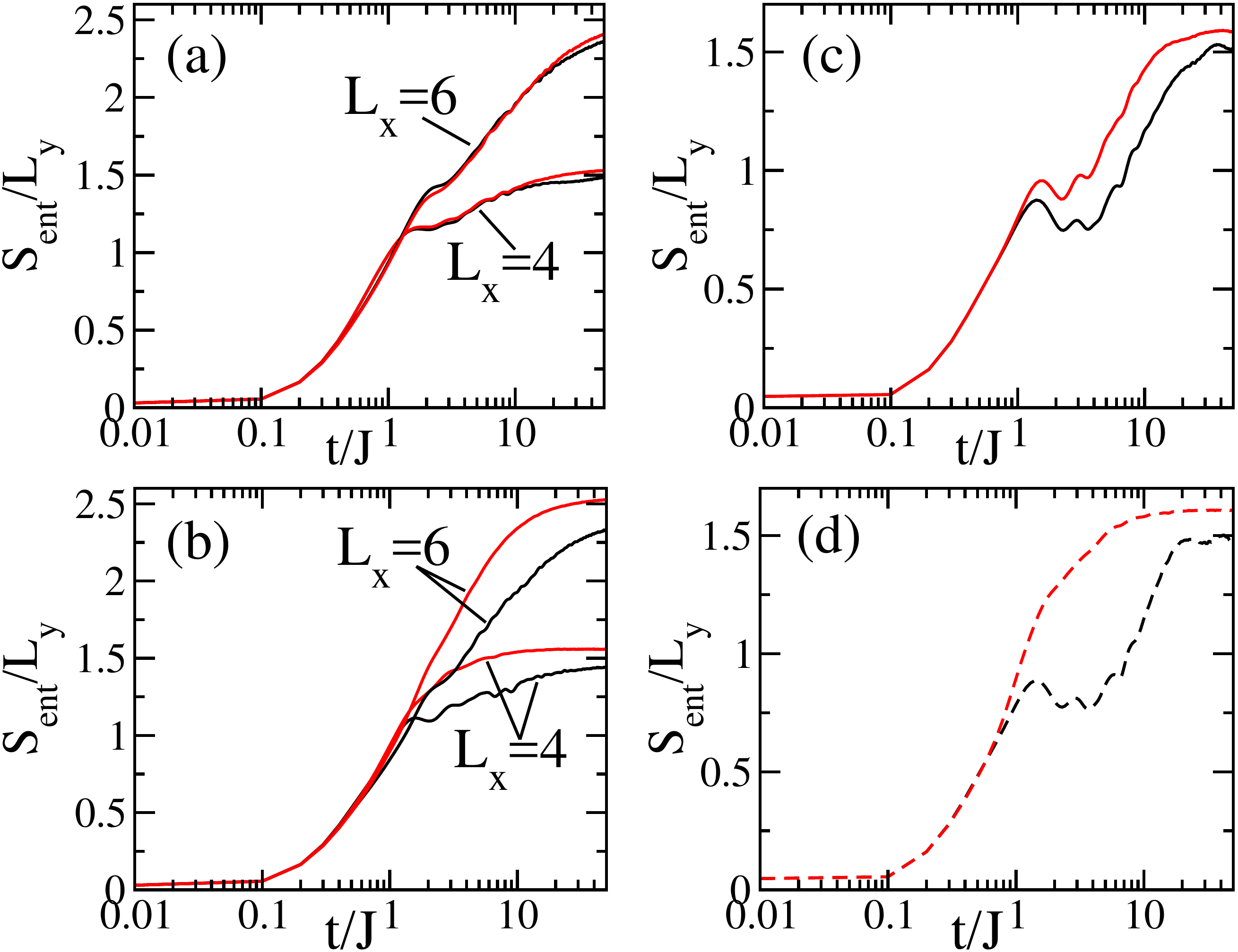}
  \caption{(Color online) Entanglement entropy per chain for
  interacting spinful fermions on a $4\times 2$ ladder with
  $n_\uparrow=n_\downarrow =2$ ($4000$ samples) and a $6\times 2$
  ladder ($400$ samples) with $n_\uparrow =4$, $n_\downarrow =2$. The
  interchain couplings are $J_\perp=0.1$ (black lines) and $J_\perp=1$
  (red lines). Left panels: rung initial state, right panels: diagonal
  initial state. (a,c) $U=1$ and (b,d) $U=6$.}
\label{Fig11}
\end{figure}

Another way of breaking the one-dimensionality of the dynamics is to
include interactions. As demonstrated in Fig.~\ref{Fig11} the
entanglement entropy then depends on the strength of the interchain
coupling $J_\perp$ even without the diagonal couplings.
For spin chains it has been shown that the entanglement entropy
increases logarithmically in the many-body localized
phase.\cite{ZnidaricProsen,BardarsonPollmann} While some of the data
in Fig.~\ref{Fig11} might be hinting at such a scaling, the system
sizes are too small to observe scaling over a large time interval. We
also note that it has recently been argued---based on numerical
data---that the entanglement growth in a Hubbard chain with potential
disorder does not grow logarithmically but rather follows a power law
with an exponent much smaller than $1$.\cite{PrelovsekBarisic}

\begin{figure}
  \includegraphics*[width=0.99\columnwidth]{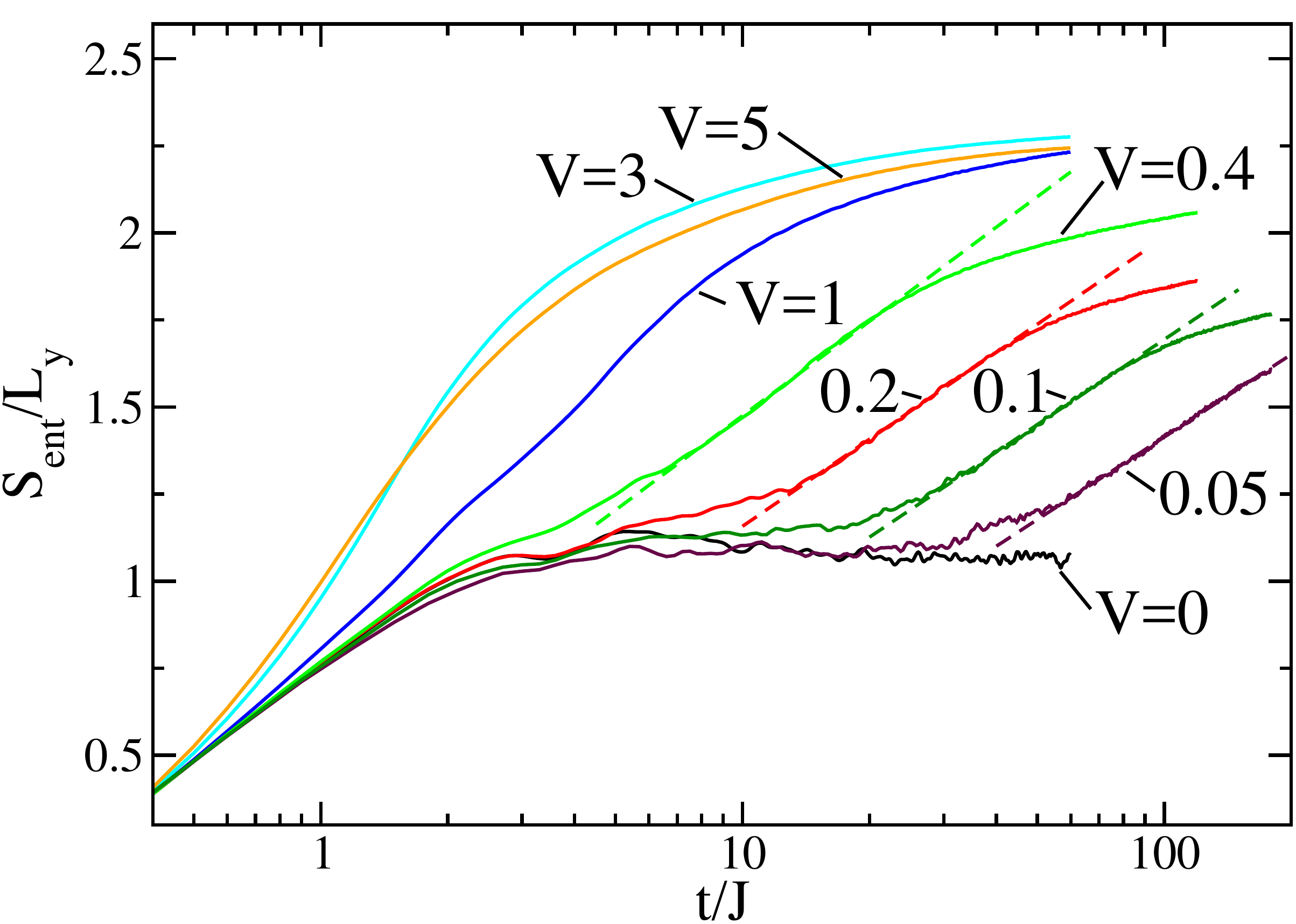}
  \caption{(Color online) Entanglement entropy per chain for
  interacting spinless fermions on a $8\times 2$ ladder with box
  disorder $D=2.5$ prepared in the rung initial state. Results for
  $1000$ samples with $J_\perp=1$ are shown. The dashed lines are
  logarithmic fits.}
\label{Fig12}
\end{figure}

In addition to the spinful case we therefore also consider the
spinless case, see Fig.~\ref{Fig12}.
In this case we do see clear signatures of a logarithmic scaling for
small interactions $V$ which seem to indicate that the ladder for
$D=2.5$ is already in the many-body localized phase. Determining the
phase diagram of the ladder as a function of disorder strength $D$ and
interaction $V$ is difficult using exact diagonalization because of
the limited system sizes accessible and is beyond the scope of this
paper.

\section{Conclusions}
\label{Conc}
We have studied non-equilibrium dynamics and localization phenomena in
fermionic Hubbard ladders with identical disorder along the chain
direction using analytical calculations in limiting cases as well as
exact diagonalizations. In the free fermion case we confirm that a
perpendicular coupling between the chains does not affect the dynamics
for an initial state where all even sites on the chains are occupied
by one fermion while all odd sites are empty (rung occupied
state). Anderson localization in the chains {\it appears to be stable}
in such a setup simply because turning on the perpendicular interchain
couplings does not affect the dynamics at all.

In order to study the differences in the response to interchain
couplings between an Anderson and a many-body localized system in a
non-trivial setup, we considered to either modify the initial state,
or to allow for additional diagonal hoppings between the chains. 

For the modified initial state---where even sites are occupied by one
fermion on even legs and odd sites on odd legs (diagonal occupied
state)---we did not find any qualitative difference between the
Anderson and the many-body localized state. In both cases interchain
coupling leads to a complete decay of the order parameter for a
two-leg ladder. At least for small systems there is also no
discernible difference in the time scales for the decay of the order
parameter between the interacting and the non-interacting model.

Similarly, we found that the order parameter for the rung occupied
state does decay also in the non-interacting case if we allow for
diagonal hoppings which truly couple the chains. Qualitatively, there
is again no difference between the Anderson and the many-body
localized case: in both cases the initial order is unstable to generic
couplings between the chains.

While a more detailed analysis of the long-time average of the order
parameter, the decay time, and of the entanglement entropy does reveal
quantitative differences between the non-interacting and the
interacting case, coupling chains with identical disorder {\it in a
generic way} does not appear to be a 'smoking gun' experiment to
distinguish Anderson and many-body localized systems.

\acknowledgments
We acknowledge support by the Natural Sciences and Engineering
Research Council (NSERC, Canada) and by the Deutsche
Forschungsgemeinschaft (DFG) via Research Unit FOR 2316. We are
grateful for the computing resources provided by Compute Canada and
Westgrid. Y.Z. thanks Prof.~J.~Cho for useful discussions. Y.Z. is
supported (in part) by the R\&D Convergence Program of NST (National
Research Council of Science and Technology) of Republic of Korea
(Grant No. CAP-15-08-KRISS).

\end{document}